# Does Light Gravitate?
# (Proposal on New Test of Equivalence Principle)

Anatoli Vankov
(Physics Department, Eastern Illinois University, cfaav@eiu.edu)

**Abstract**

On the basis of the relativistic mass-energy concept we found that a proper mass of a test particle in a gravitational field depends on a potential energy, hence, a freely falling particle has a varying proper mass. Consequently, a multitude of freely falling reference frames cannot be regarded as a multitude of equivalent inertial reference frames. There is a class of experiments, which allow distinguishing between them. If so, a demonstration of a violation of the Equivalence Principle is possible. It is shown that a variant of the classical Pound-Rebka-Snider experiment on a photon frequency shift in a gravitational field, if conducted in a freely falling laboratory, would be such a test.

Abbreviation: SRT- the Special Relativity Theory,
GRT- the General Relativity Theory.
EP   - the Equivalence Principle,
PRS - the Pound-Rebka-Snider (experiment)

*Introduction*

The phenomenon of gravity is known to be incompatible with the Special Relativity Theory (SRT), so the General Relativity Theory (GRT) is currently considered the only reasonable field theory of gravity. However, attempts to develop a Relativistic Quantum Gravitodynamics failed. A general feeling about such a situation may be expressed as the expectation of a breakthrough by a discovery of a radically new fundamental physical principle governing the behavior of matter under both relativistic and quantum-mechanical conditions. Then known problems in the Standard Models of Particle Physics and Cosmology would possibly be understood.

It is thought less probable but not to be excluded that a breakthrough could happen by finding out that the GRT is inherently inconsistent. As a mater of fact there are a few direct tests of the GRT under "weak field" conditions and none for "strong field". Any evidence of a violation of the Equivalence Principle would be a signal of such inconsistency, that is, an inherent contradiction. In other words, the GRT being a falsifiable theory could be discarded in principle by a properly formulated observational test revealing the contradiction. In reality a situation is not as simple as that because effects of such significance are practically very small and hard to observe with the required precision. Besides, plenty of room is left for model corrections. Such difficulties are seen, for example, in NSF gravitational experimentation programs.

One of the "classical tests" of the GRT was carried out of weighing a photon in the gravitational field (the experiment by R.W.Pound and G.A.Rebka [1], R.W.Pound and J.L.Snider [2], the PRS experiment, for abbreviation). Though the measured effect is extremely small the experiment has been eventually conducted with a high precision (about one percent), and the result was in a good agreement with the GRT prediction. By that time the Equivalence Principle (EP) was firmly justified in different measurements, and the PRS experiment was considered as additional evidence supporting the EP. The GRT predicts the observed photon frequency shift proportional to the strength of a gravitational field. According to the EP the shift disappears as the field disappears in a freely falling frame carrying both a photon resonance emitter and a corresponding detector. Though the experiment was not reproduced under free fall conditions due to obvious technical complications there is no doubt among the GRT community that it would result in a zero shift. But we doubt such a result for some reasons explained below. Therefore, we expect a violation of the EP and propose a repetition of the PRS experiment in a freely falling frame.

## 1. Current status of the GRT

It is generally acknowledged that the EP is the physical basis of the GRT. In this sense, the GRT might be considered to be more "general" than the SRT. But some GRT experts argue [3,4] that there is no need to refer to the EP while formulating the GRT because the EP mathematically reflects a trivial fact of possibility to locally approximate a curved 4-space by a flat space metric. However, the physical meaning of the EP seems to go beyond the mathematical treatment. The EP physical formulation is as follows. *In an arbitrary gravitational field no local experiment can distinguish a freely falling non-rotating system (local inertial frame) from a uniformly moving system in the absence of a gravitational field.* There is another formulation of the EP: *at every point of spacetime it is possible to choose local coordinates so that all physical laws take the same form as they are formulated in the SRT in the absence of gravity.*

As was shown (see, for example [5]) the EP rests on the equality of the gravitational and inertial mass, hence, the equality is postulated in rather than deduced from the GRT. At the same time the GRT has an ability to predict some controversial *Machian effects* (frame dragging, in particular). Those effects, in fact, show the way of detecting a source of the "disappearing" gravity field in a free fall state. There was a troubling moment in the GRT history in connection with a necessity of accepting a local energy disappearance and introducing a pseudo-tensor for the stress-energy-momentum term. Though the pseudo-tensor still looks much as a foreign body in the theory, one may argue that it is not a drawback of the theory but rather that nature behaves accordingly. Among disputable issues of modern Physics the most important one is a failure of the GRT to be subject to renormalization and quantization. It makes the current situation in the gravitational theory quite problematic and very challenging. In this situation we are going to raise the old question of physical validity of the EP. The starting point will be the relativistic phenomenon of a proper mass variability, which follows from the SRT.

## *2. A proper mass variability in the SRT mass-energy concept*

Let us consider a point-like particle of mass *m* in a central gravitational force field with a potential $\phi(x)$ produced by a solid sphere of mass $M_R \gg m$ and radius *R*:

$$\phi(x) = -GM_R/x \qquad (x \geq R) \qquad (1)$$

where *G* - the gravitational constant, *x* - the distance between the center of the sphere and the particle. The particle can be slowly moved along the radial direction with use of some ideal transporting device supplied with energy. The whole system is isolated. In accordance with the SRT mass-energy concept a change of potential energy of the particle should be equal to the corresponding change of its proper (rest) mass:

$$dm = [GM\, m(x)/c^2 x^2]\, dx, \quad (x \geq R) \qquad (2)$$

In this conceptual example the question about work-energy balance (does the particle perform work?) is viewed differently in the SRT and the GRT. In the latter a proper mass is assumed to be constant while a field provides the particle with energy (the field works on the particle). Then the controversy arises around the question of a separation of energy of a field and a total mass-energy of the system. In the SRT there is no room for the field energy additional to the total mass-energy of interacting particles. In the static case of a conservative force field a transporting (supporting) device with an energy source should be included in the system. Due to an energy exchange between the particle and the transporting device the particle turns out to be bound. The total energy of the particle under consideration is characterized by two varying components: the proper mass-energy and the binding energy, the sum of them being constant.

The above difference between the SRT and the GRT is of principle importance. Both theories claim to be competent to consider the conceptual example. They look apparently self-consistent but result in different Physics. Bearing this warning in mind we will continue to consider the conceptual example on the SRT basis.

From (2) the proper mass is found as a function of the distance:

$$m(x) = m \exp(-x_R / x), \quad (x \geq R) \qquad (3)$$

or in a "weak field" approximation:

$$m(x) = m(1 - x_R / x), \quad (x \gg x_R) \tag{3a}$$

where $x_R = GM/c^2$. At $x = x_R$ in (6) the force is maximal, while the particle tends to acquire a minimal potential energy on the surface of the sphere. It means that the particle approaching the center of gravity performs work against a reaction force caused by a gravitational force. Therefore, the proper mass of the particle has to be reduced by the equivalent amount of energy to be given to the transporting device. A binding energy (a mass defect) should be interpreted as a potential energy change.

In the general case of a multi-particle isolated system, the total mass defect is shared by interacting particles, an individual share being determined by a particle mass distribution and a final state of the system. The behavior of the system seems to be governed by the principle: the *proper mass of the system tends to minimum*. Using the same approach one can see, for example, that in the case of two identical point-like gravitating particles a proper mass should be a function of the distance $x$ between them:

$$m(x) = m/(1 + x_0 / x) \tag{4}$$

where $m$, as before, is the maximal proper mass at infinity, and $x_0 = Gm/c^2$. Notice that we again use the model of an isolated system of particles moved with use of a set of ideal transporting devices as a part of the system. It allows us to consider properties of the static potential in terms of the SRT concept of mass-energy conservation.

It is seen that the static potential $\phi(x) \sim 1/x$ in fact describes an asymptotic behavior of interacting particles at a large distance comparing with $x_0$. In the limit as $x \to 0$ a proper mass vanishes as a result of the "energy exhaustion" effect. When a proper mass variability (4) is taken into account a gravitational force between two identical point-like particles takes a form of

$$F(x) = mc^2 x_0 /(x + x_0)^2 \qquad (5)$$

The "exhaustion" effect means that a singularity for a point-like source (a classical "self-energy" divergence) turns out to be eliminated in agreement with the SRT mass-energy concept. Notice that a "field' in the GRT is inexhaustible.

In the case of the potential generated by a solid gravitating sphere (1) a force exerted on a test particle along with its proper mass should vanish on the surface of the sphere with a sufficiently big ratio $M_R / R$ (say, in the gravitational collapse case)

$$F(x) = (mc^2 x_R / x^2) \exp(-x_R / x), \qquad (R < x < x_R) \qquad (6)$$

At $x = x_R$ a force is maximal, therefore, in the region ($R < x < x_R$) the particle is "locked": the force increases with a distance (confinement effect).

In accordance with (6) the potential energy function in general has a form:

$$m\phi(x) = \int_x^\infty F(x)\, dx = -mc^2 [1 - \exp(-x_R / x)], \quad (x \geq R) \qquad (6a)$$

where $m$ is the proper mass at infinity, and $x_R = GM_R / c^2$, as in (3). In (6) and (6a) we have a relativistic form of a classical gravitational force and potential function for a massive sphere with critical parameters $x_R$ and $x_0$. Obviously, the Newtonian limit takes place at $x_R \ll x$.

The phenomenon of a proper mass variability being a consequence of the general SRT mass-energy concept takes place in any type of interaction. Nuclear energy is the commonly known example. Let us consider the effect in the case of the Coulomb force. Assuming that an electric charge is not affected by a proper mass variation (in accordance with observations) one can find the effect due to particle-antiparticle electric attractive force:

$$dm = (kq^2/c^2 x^2)dx \qquad (7)$$

where $k$ is the electric constant in the Coulomb law, and $q$ is the electric charge. Then

$$m(x) = m(1 - x_a/x), \qquad (x \geq x_a) \qquad (8)$$

A proper mass vanishes at $x_a = kq^2/mc^2$. It means that at this distance "something should happen", apparently, annihilation. At the "annihilation distance" $x_a$ the Coulomb potential energy turns out to be equal to the proper mass energy equivalent, which has to be converted into electromagnetic energy.

As is seen, both the gravitational and the Coulomb potential have the same source of energy (a proper mass). This result casts a light on the problem of a source of energy in a static electric field. One has to conclude, for example, that the energy stored in a charged capacitor is due to its increased proper mass as a result of work performed by an external inertial force (a charged sphere has to be heavier). Existing electromagnetic theory does not reveal this connection of gravity and electricity.

Obviously, in the case of a repulsive force between charge-like particles a proper mass change is opposite to that for unlike charge particles. The effect is appreciable at distances comparable with $x_a$, and indefinitely increases when $x \to 0$:

$$m(x) = m\,(1 + x_a / x) \tag{9}$$

Remember that we are restricted to the SRT assumption of point-like particles, hence, the above critical distances so far should be treated as parameters of the point-like particle model.

In Newtonian Physics the force-energy theorem and the mechanical energy conservation law are formulated without explanation of the physical meaning of potential or kinetic energy. The GRT is not very helpful either in this respect. The SRT mass-energy concept makes terms defined on the relativistic basis of the total energy conservation law in consistence with Newtonian Physics. The relative effect of a proper mass variation $dm/m$ is extremely small under Earth or Sun conditions. Hence the static potential in a multi-body system may be found in the "weak field" approximation by integration over sources in the manner of the Coulomb potential. However, *a classical gravitational field in principle turns out to be non-linear* (non-linearity would undoubtedly be important in astrophysical applications).

Further we use the term "gravitational field" in a sense of a gravitational force field. We do not exploit the SRT Dynamics equations of motion in a gravitational field because those equations, if treated in terms of field, were allegedly found contradictory to observations. The situation looks quite controversial and should be further investigated.

### *3. Is the SRT really inconsistent with observations?*

The above question tacitly expresses the conviction that the *GRT is consistent* with observations. However, the GRT field presentation does not seem fully satisfactory because it does not explain the physical nature of the concept of "a field as a carrier of gravitating energy as a source of the field" as an addition to the EP. Next examples illustrate controversy in this concept.

a. Consider annihilation of slow electron and positron pair. Observation of this process shows the exact SRT mass-energy balance with no energy gained *from the field* in addition to the energy equivalent of the initial proper mass.

b. According to the GRT a kinetic mass is the one of a gravitating type, hence, an assessment of the main cosmological parameter (the critical mass density, which determines a universe gravity pull) should include a kinetic mass of both massive matter and massless particles. In fact, the assessment is based on counting a proper (rest) mass only. Kinetic energy depends on a reference frame choice, hence, a gravitational force cannot be affected by it.

c. The GRT predicts gravitational and electromagnetic radiation of an accelerating particle or a body, in a free fall state, in particular. For example, a gravitational radiation from binary stars is searched by means of energy balance counting. However, had the radiation been unambiguously found it could have manifested a violation of the EP.

Main references on attempts to develop a gravity field in a flat (Minkowski) space are given in [6]. The conclusion was made that the SRT is not compatible with gravity. For example, a scalar field was ruled out because it does not couple gravity to light. As was emphasized, in the GRT a freely falling particle gains energy from the field, and the same is true for a photon. Therefore, gravitational properties of a photon should be the problem one has especially to be concerned with. An apparently strong GRT position in this issue is the statement that *all forms of energy are subject to gravitation,* electromagnetic waves (light) included. The statement in fact has a status of a postulate in addition to the EP. Both the GRT and the SRT, having no quantum-mechanical extension in a gravity theory, cannot provide a theoretical basis for gravitational properties of massless particles. In practice both are guided by circumstantial arguments.

The SRT gives a picture of a freely falling particle, which spends its own source of energy (a proper mass) to gain kinetic energy. A physical nature of the kinetic energy is not clear at this level. Evidently, we need a quantum-mechanical interpretation of an interaction of a particle with the "physical vacuum". Further we will see that regardless of the quantum-mechanical properties of the kinetic mass its dynamical role in the presence of gravitational and inertial forces can be well understood in the SRT. As was discussed above, one should expect that the kinetic mass does not gravitate. Similarly, a photon having no proper mass is not expected to change its energy in a gravitational field. Then how to interpret the PRS experiment, which showed a photon frequency shift?

According to the GRT the shift is due to the change of energy (frequency) of a *gravitating photon* in a gravitational field. The shift will disappear if measured in free fall because the field disappears. We suggest another picture. A *non-gravitating photon* does not change its frequency in flight. This is a difference in resonant frequency of an emitter and a detector placed at different equipotential surfaces that causes the observed effect. In other words, the resonant shift is due to a difference in a proper mass of nuclei of the emitter and the detector. Which interpretation of the experiment is true may be and should be verified in some crucial test. In the following section we show that the *PRS experiment in a freely falling frame* would be such a test. We predict that in this variant the observable frequency shift would be the same as that measured in the original PRS stationary experiment. A confirmation of this statement would mean a violation of the EP. So, the proposed test is vital in the problem of compatibility of the SRT with a gravitational theory. The problem is considered next in more detail.

### 4. Brief review of the SRT Mechanics

The SRT Mechanics is generally known as the Relativistic Kinematics of a free particle considered in some "inertial" reference frame. To deal with a process of transition from one frame to another we need to use the SRT dynamical equations, which are invariant under Lorentz transformation:

$$\frac{d}{d\tau}[m(\tau)\frac{dx^\alpha}{d\tau}] = K^\alpha, \quad (\alpha = 1,2,3,4) \tag{10}$$

They describe a particle motion on some world line $x^\alpha(\tau)$ in an inertial reference frame with Minkowski coordinates $x^\alpha$ and a 4-velocity $\frac{dx^\alpha}{d\tau}$, where $K^\alpha(\tau)$ is a Minkowski 4-force vector, and $d\tau$ is a line arc-length. The equations (10) have been found as a generalization of Newtonian Mechanics by means of the Lagrangian formulation of Relativistic Mechanics. They show that the rate of 4-momentum change equals the Minkovski force. By definition of a time-like world line of a moving particle (with a speed less than the speed of light $c$) we have the fifth equation:

$$\frac{dx^\alpha}{d\tau}\frac{dx^\alpha}{d\tau} = -1 \qquad (11)$$

which makes the problem definite with respect to five unknown functions $x^\alpha(\tau)$, $m(\tau)$. As was emphasized by Synge [3], a proper mass variation along the world line is explicitly seen from the next equation obtained from (10) and (11):

$$\frac{dm}{d\tau}\frac{dx^\alpha}{d\tau} + m\frac{d^2 x^\alpha}{d\tau^2} = K^\alpha \qquad (12)$$

Remember that a photon is a massless particle, hence for a photon

$\frac{dx^\alpha}{d\tau}\frac{dx^\alpha}{d\tau} = 0$ in a flat space having the Minkowski metric $\eta_{\alpha\beta}$

$$d\tau^2 = -\eta_{\alpha\beta} dx^\alpha dx^\beta = -dx^\alpha dx_\alpha \qquad (13)$$

A free particle has a constant proper mass, hence, the equation of motion of a free particle is

$$\frac{d^2 x^\alpha}{d^2\tau} = 0 \qquad (14a)$$

In a non-inertial reference frame one may use a general coordinate system $[x^\alpha(x^\beta)]$. Then equation (14a) becomes

$$\frac{d^2 x^\alpha}{d\tau^2} + \Gamma^\alpha_{\beta\gamma}\frac{dx^\beta}{d\tau}\frac{dx^\gamma}{d\tau} = 0 \qquad (14b)$$

where $\Gamma^{\alpha}_{\beta\gamma}$ is the metric connection of $g_{\alpha\beta}$:

$$d\tau^2 = -g_{\alpha\beta} dx^{\alpha} dx^{\beta} \tag{15}$$

So far, nothing has happened: we have the same particle in the same state of free motion, described in an arbitrarily chosen coordinate system. Formally the flat metric $g_{\alpha\beta}$ is not the Minkowski one $\eta_{\alpha\beta}$ but this difference is a matter of pure mathematical treatment. A physical interpretation begins with the introduction of the EP in the presence of a force field. The EP requires an appropriate form of $\Gamma^{\alpha}_{\beta\gamma}$ being a metric connection of a *non-flat metric* $g_{\alpha\beta}$ in the presence of both gravitational and inertial forces. The intrinsic curvature of spacetime is now introduced, consequently, $\Gamma^{\alpha}_{\beta\gamma}$, and $g_{\alpha\beta}$ should be interpreted in terms of a force field. At this point the GRT has come in conflict with the SRT because a proper mass of a particle in free fall should depend on a gravitational potential to comply with the energy conservation law in its SRT form. We will see that dynamical properties of a relativistic mass are ignored in the EP. Remember that the EP basically rests on the postulate of the equality of gravitational and inertial mass, but a formulation of the postulate loses a physical sense in the Relativistic Dynamics.

For the sake of convenience one may come from the description in spacetime ($\alpha = 1,2,3,4$) to the description in 3-space ($i = 1,2,3$) and time $t$ ($\alpha = 4$) using the relation $d\tau = c\,dt/\gamma$ and introducing relative ("ordinary") forces $F_i$:

$$F^i = c^2 K^i / \gamma \tag{16}$$

Now the equations of motion take the form:

$$\frac{d}{dt}(m\gamma u^i) = F^i \tag{17a}$$

$$\frac{d}{dt}(m\gamma c^2) = F^i u_i \tag{17b}$$

where $u^i(t) = dx^i/dt$ ($i = 1,2,3$) is the relative 3-velocity, and the proper mass $m$ is time dependent. The second equation reflects the energy conservation law in the relativistic form. Examination of the above equations reveals a different role of a relativistic mass in a change of energy and momentum under different conditions. Without solving the equations we may draw some important qualitative conclusions consistent with the relativistic balance of mass-energy and momentum.

Obviously, any constant force exerted on a massive particle causes a change of momentum. A kinetic mass-energy changes if the force is not perpendicular to the velocity. If it is perpendicular, the momentum changes direction (but not magnitude) while the kinetic and total mass-energy remain constant and the particle is kept bound. *Its proper "bound" mass is less than that at infinity.* This is true for the Kepler circular motion, in particular. If the particle is in a state of free fall from infinity (initially being at rest) the total mass-energy $m_t$ (the sum of varying both proper $m_p$ and kinetic $m_k$ mass) equals the proper mass $m$ of a free particle at infinity:

$$m_t = m_k(t) + m_p(t) = \gamma(t) m_p(t) = m , \quad (m_p(t) \leq m) \tag{18a}$$

It would be different in the case of uniform acceleration of a particle due to a constant inertial force. A particle continually gains kinetic energy from an external energy source while *a proper ("bound") mass is being constant but bigger than that at infinity:*

$$m_k(t) + m_p = \gamma(t) m_p = m(t), \qquad (m_p \geq m) \tag{18b}$$

where in both cases $\gamma(t)$ is a varying Lorentz factor.

In practice, the inertial force has to be initially developed over a time interval $\Delta t_i$ when the force rises. A proper mass rises synchronously and reaches a maximum gain $\Delta m = m_p - m$ in (18b) when both the force and the proper mass became constant. During a period of uniform acceleration the kinetic mass continually increases. A force pulse would be completed when the force drops down to zero over a final time interval $\Delta t_f$. Correspondingly, $\Delta m$ comes to zero, and the particle will be in a state of uniform motion with the mass-energy balance described by the known formula in the SRT Kinematics:

$$m_t = m_k + m = \gamma m \tag{18c}$$

It is seen that *the proper mass pulse $\Delta m$ replicates the force pulse and determines a direction of a mass-energy current between a test particle and its interacting partners. A magnitude of the current is proportional to the proper mass difference, that is, the potential difference developed due to a force change.* Therefore, a third derivative of coordinate should be generally specified in practical applications of the SRT Dynamics. An example of a "forbidden" for the SRT problem is a periheleon precession in the Kepler motion. The observation of the effect played a role of one of the seemingly successful tests of the GRT. Certainly, an exact solution of this problem could be found in the SRT Dynamics had the SRT compatibility with gravity been established. In this work we are more concerned with the previously raised question about the role of a kinetic mass in a gravitational force. The question may be formulated as next: does light gravitate? We are now prepared enough to clear up this issue.

The above examples show that gravitational and inertial masses are not equivalent in many ways. For example, a particle in a state of free fall gains its kinetic mass and a momentum at expense of a proper mass reduction, that is, differently than in an inertial force field. Mutually, a kinetic mass is differently related to forces of different types. The emerging kinetic mass will be eventually materialized, but before that it does not contribute to the gravity force. Suppose an observer in a freely falling frame is allowed to communicate with an outside laboratory and able to register his position with respect to the attractive center. Then, in accordance with the equations (10-12), he would be able to conclude that a proper mass decreases due to a gravitational pull with no role of a kinetic mass. As the second example (see the discussion in the previous section) consider a gravitational force between two massive particles moving in parallel. It should be determined by a proper mass of the particles, otherwise, it would depend on an arbitrary choice of a reference frame. The conclusion drawn from the SRT approach is that the kinetic mass does not gravitate, only the proper mass does.

The crucial question remains whether the observer in a freely falling frame is able to detect the presence of a gravitational center *not communicating with the outside world* (in a violation of the EP). Before answering this question we need to finish considering the role of the kinetic mass in a momentum change.

As is seen from the SRT Dynamics equations, a pulse of any force changes a momentum *by action on a total mass* $\gamma m$. A photon, the total mass of which equals the kinetic one, may be considered as a particle in a limit as $m \to 0$ and $\gamma m = const$. Considering a motion of such a "photon-in-limit" in gravitational field one has to conclude that the result of a force pulse should be a change of momentum but not energy (frequency). Otherwise the equations would not comply with the total energy conservation law stated in the SRT form.

This topic has a quantum-mechanical aspect. A photon frequency $\upsilon_{ph}$ as a characteristic of a total mass-energy is both a relativistic and a quantum-mechanical feature given in the form

$$h\upsilon_{ph} = m_{ph} c^2 \qquad (19a)$$

where $m_{ph}$ is a photon kinetic mass-energy, which is equal to a total (relative) mass of a photon. According to the Louis De Broglie concept of a wave nature of matter [7] any particle should be considered in a "physical vacuum" an oscillator having a proper frequency $\upsilon_{pa}$ similarly related to the proper mass $m_{pa}$:

$$h\upsilon_{pa} = m_{pa} c^2 \qquad (19b)$$

On this basis he found a dispersion relation and a wavelength scale $\lambda(p)$ for a particle interfering with matter:

$$\lambda = h/p \qquad (20)$$

where $p$ is a momentum.

It is seen that a proper mass variation in a gravitational field is a property, which characterizes a variable energy scale of a material system in a gravitational field. Resonance energy (frequency) of a photon emitted by an excited nucleon depends on a proper mass of a nucleon, that is, its potential energy. At the same time the resonance frequency of a photon in flight should be constant in a conservative (gravitational) field. We have to conclude this section with the statement that according to the SRT mass-energy concept a photon in a gravitational field does not change its energy (frequency). In other words, light does not gravitate. The argument that the SRT is not compatible with gravity because a field representation in a flat space resulted in no coupling of light to gravity might be wrong if light does not gravitate in reality. To clear up the situation we are going to carefully analyze two classical GRT tests involving light: bending of light and "red-shift" of a photon in a gravitational field.

## 5. *The light bending phenomenon*

The light bending phenomenon is directly related to the question of the change of a photon's momentum. We need to realize why and how a photon could change direction without changing its total energy (frequency). In consistence with the equations (17), the change of direction is due to the gravitational force pulse. To keep a frequency constant the photon needs to change the magnitude of velocity. We come to the striking conclusion that a photon crossing equipotential surfaces needs to vary the speed of light. The wavelength changes accordingly. In other words, a light interference with gravity occurs through a change of momentum (not energy). A special form of refraction takes place: photon speed decreases with a decrease of potential energy (and vice versa), frequency being constant. It could be described in terms of a photon entering a medium with a changing "optical" density as compared to the "free space".

To estimate the gravitational refraction effect one may consider a system of two identical bodies in the form of charged spheres of equal masses $m$ and equal like charges $q$. We have $Gm^2 = kq^2$ when the system is in equilibrium being far away from a gravitational center. Remember that $k = 1/4\pi\varepsilon_0$ where $\varepsilon_0$ is the permittivity of free space. The distance between spheres is fixed in a laboratory frame. When the system is in a state of free fall onto a gravitational center, the distance is observed unchanged (otherwise energy will arise "from nowhere"). An observer in the freely falling laboratory is unable to notice a change of energy scaling (a proper mass decreasing) without communication with the outside world. For him the system of two bodies is kept in equilibrium. The EP in this respect is perfectly valid. On the other hand, an outside observer is able to distinguish between total and kinetic masses and conclude that the system should be kept in equilibrium in spite of a proper mass decrease. It is possible only if a decrease of the gravitational force between two spheres would be accompanied by a Coulomb force decrease at the same rate. In other words, a permittivity of space in the presence of a gravity field should increase. The outside observer will find the effect looking at (6) and (6a) and bearing in mind that an electric charge is not affected by gravity. Then a permittivity dependence on gravitational field strength may be easily found:

$$\varepsilon(x) = \varepsilon \exp(2x_R/x) = 1/[1+\phi(x)/c^2]^2 , \quad (x \geq R) \quad (21a)$$

where, as before in (3), $x_R = GM/c^2$, $c$ and $\varepsilon$ are *the speed of light and the permitivity in free space,* correspondingly. It would be premature to discuss the effect on the "Schwarzschild surface" $x_{Sc} = 2x_R$ or below it. Under the "weak field" condition we have

$$\varepsilon(x)/\varepsilon \cong 1 + \frac{\phi(x)}{c^2} \cong 1 + \frac{2x_R}{x} , \quad (x \gg x_R) \quad (21b)$$

We have to conclude that the outside observer experimenting with light should find that a photon is slowing down when it crosses equipotential surfaces "down", and speeding up when traveling "up", its frequency being kept constant. Now the speed of light $c(x)$ depends on a coordinate of an equipotential surface:

$$i_g(x) = c(x)/c = 1 - \frac{2\phi(x)}{c^2} \quad (22a)$$

We may call the ratio $i_g$ the gravitational index of refraction. Under "weak field" conditions it takes the form

$$i_g(x) = c(x)/c \cong \sqrt{1 - 2x_R/x} \cong 1 - x_R/x , \quad (x \gg x_R) \quad (22b)$$

One has to realize that once a photon is emitted along an equipotential plane in a uniform gravitational field, the photon does not change its direction until gravitational potential changes. As was discussed before, a pulse of force is needed to change a momentum by affecting a kinetic mass. It means that there is indeed no coupling of a photon to gravity. Back to the time before the GRT has been completed the question of variation of speed of light in a gravitational field was raised by Albert Einstein and others [7]. This idea was shortly abandoned due to its inconsistency with the GRT in its final form, in which a kinetic mass of a photon has the same gravitational properties as any massive particle.

Assessment of the bending angle of light passing the Sun in the SRT approach is numerically close to that in the GRT. In the first approximation the angle $\delta$ is given by the formula similar to that in the GRT:

$$\delta = 4M_S G / c^2 R_S \qquad (23)$$

where $M_S$ and $R_S$ are correspondingly the proper mass and the radius of the Sun measured by outside observer at rest. We know that $M_S$ is a "bound mass", binding energy being a potential energy of the system. In the GRT approach $M_S$ is interpreted in terms of field and is called a "total mass of the spherically symmetric solution". Numerical difference (SRT versus GRT) in $M_S$ grows with a "field strength" increase. As is known, the angle assessment initially made by Albert Einstein was twice as low as given in (23). Later, the magnitude was doubled to make the assessment consistent with the Schwatzcshild metric characterized by $g_{tt} = -(1 - 2GM / c^2 r)$. The factor 2 in $g_{tt}$ is responsible for the GRT effect of time dialation due to gravity.

A singularity on the Schwatzcshild sphere of the radius $r_{Sc} = 2GM / c^2 R$ was the subject to numerous discussions. As was shown, in the SRT concept the singularity is excluded due to an exponential form in (3), consequently, in (6) and (6a) of the potential of a massive sphere. The SRT interpretation of the effect of time dialation is different: an outside observer detects a decrease of the speed of light in the gravitational field, therefore, a communication signal is delayed.

There was a discussion whether the GRT result (23) is really consistent with the EP. This question has a physical sense, because the formula derived purely from the EP is exactly what Albert Einstein initially found (with factor 2 instead of 4 in (23), see, for example, [8]). Additional factor 2 comes beyond the EP due to *assumed* gravitating properties of light, as we previously emphasized. The formula (23) ensures an energy conservation law *as it is treated in the GRT field terms*. The reason for the controversy is that the EP rests on the postulated equality of gravitational and inertial mass. The postulate had been formulated before the relativistic theory, and nobody knows what it says exactly if applied to the light.

### *6. A frequency shift of a photon in a gravitational field*

All things considered, our interpretation of the PRS experiment is as follows. For measuring a photon frequency shift both a resonance emitter and a detector should be put at rest at a laboratory frame. Identical nuclei of an emitter and the detector are placed at different equipotential surfaces. Therefore, they have necessarily different proper masses. It is important to use the solid material in both the emitter and detector with the same atomic structure to eliminate a difference in additional binding energy due to atomic structure. Then a proper mass difference will be exclusively due to the gravitational potential difference. Thus, the proper mass determines energy scaling of matter on the equipotential surface. This is the cause of a corresponding shift in a resonance frequency of electromagnetic transition in excited nuclei. We have to emphasize again that an emitted photon being in flight does not change its frequency because its proper mass is zero. If an emitter is placed below a detector the latter will register an apparently red-shifted photon due to the difference in resonance frequencies of the emitter and the detector. After linearization of (3) at $R = R(Earth) >> x_R$ with $g = 9.8 m/s^2$ and making use of (19b) we have a relationship between the change of a proper mass of nuclei $\Delta m$, the distance between equipotential levels $\Delta r$, and the difference in a resonance frequency of the emitter and the detector under Earth condition of a uniform gravitational field:

$$h\Delta f = c^2 \Delta m = mg\Delta r \tag{24}$$

Now let us discuss the effect of changing speed of light. An in-flight increase of wavelength of a photon traveling "upward" due to its speeding up may be called a "red shift". However, the detector is "tuned" exactly on the wavelength of the incident photon in the moment of detecting. Hence there is not any "red shift" at all in resonant wavelength spectra of the detector and incident photons in the moment of absorption. As was emphasized before, resonance frequencies do not match, and for this reason experimenters detect an apparent "red shift". In general, photons travelling in space in the presence of a gravitational field are distinguished by their resonance energy (proper frequency). Photons with the same resonance frequency would have the same wavelength if and only if they belong to the same equipotential surface.

Interestingly enough, one can find the same formulae (24) and the same argument of energy conservation law, as concerns a gravitational red shift theory, in textbooks on the General Relativity Theory. But the interpretation is different: $\Delta m$ and $m$ are referred now to the kinetic mass of a photon (19a). As opposed to the SRT picture, the photon interacts with a gravitational field and gains energy from it while a nucleon proper mass is kept constant.

*What do we expect from the PRS experiment conducted in a freely falling laboratory? In accordance with the SRT mass-energy concept, there is a difference in a proper mass and, correspondingly, in a proper resonance frequency of nuclei of an emitter and a detector placed at different equipotential surfaces. The difference takes place regardless of the reference frame, hence, is not affected by the state of free fall of the laboratory. Because a photon being emitted by the emitter does not change its frequency in flight an apparent "red shift" would be observed in the freely falling laboratory the same as in the laboratory at rest.*

### *Discussion*

We found that the SRT interpretation of the bending light phenomenon and the PRS stationary experiment differs from that in the GRT, but numerical assessments are close as far as the effects have an "weak field" order. Predictions of "strong field" phenomena would be numerically different. On the other hand, predictions of observations, which are sensitive to gravitating properties of light and made by an observer *in a freely falling frame*, are numerically different regardless of the field strength. In this case we expect a violation of the EP from the SRT standpoint. As was discussed in the fifth section, the observation of the bending of light if to be considered the EP test is not complete because the GRT time dialation effect was not measured. The same situation occurred in the case of the "red-shift" EP test. This second test cannot be considered as completed until the PRS experiment is performed in a freely falling laboratory.

Suppose the PRS experiment, if conducted in a freely falling laboratory, showed the frequency shift unchanged as compared to the stationary condition. Then the status of the SRT in a gravitational theory should be reconsidered. What then could be the prospect for a gravitational theory development? In such a situation one has to deal with the new gravitational properties of light such as the absence of coupling to gravity and variation of the speed of light in a gravitational field. Then the Lagrangian formulation of the SRT Mechanics including gravity seems to be possible. It means a principal possibility of a gravitational field representation in the Minkovski space. Obviously, it would be a non-linear gauge theory ensuring the SRT mass-energy concept of energy conservation. The total energy should correspond to the total proper mass of a system when the system is decomposed into elementary particles moved to infinity. At this point we have come to the limitation of the SRT (point-like particle model) to coup with in a way as Relativistic Quantum Electrodynamics does with point-like charges.

The SRT approach casts a new light on this problem: a singularity of the point-like source turned out to be eliminated due to limited particle energy. Both gravitational and electromagnetic forces seem to arise due to the same source of energy, that is, a proper mass of interacting particles. If so, the SRT mass-energy concept should be considered the basis of a future program of the unification of electromagnetism and gravity.

A trivial conclusion from this work is that the Lorentz invariance should be replaced by a more general principle. To clear the problem let us consider a uniform gravitational field in a space within a spherical layer $\Delta r$ around a massive sphere $r >> R$. We found in the SRT approach that every point in the space is labeled by a potential $\phi(x)$, which influences a flat space metric by a scaling factor $i_g = c/c(\phi)$, (or the gravitational index of refraction). Freely falling frames *inside a spherical layer* should be considered equivalent inertial frames if a change of the scaling factor could be ignored (depending on the formulation of a physical experiment). In such a class of experiments with a fixed scaling factor the EP is valid, and a freely falling observer finds himself in a flat space, which is seen by an outside observer as a "$\phi$-labeled" Minkowski space { $x^\alpha, \tau(\phi)$ }. The outside observer is able to discriminate between $\phi$-worlds by detecting a continually varying scaling factor $i_g(\phi)$. The factor determines an inner Minkowski $\phi$-metric characterized by the speed of light and the proper mass-energy-frequency, both reduced by the factor $i_g(\phi)$. Therefore, the outside observer detects apparent effects of a time dialation and a length contraction (a decrease of $\phi$-photon resonant wavelength (19a)). A validity of the EP means that there is no exchange of information between observers from different $\phi$-worlds. In this sense the EP is two-dimensional. Now we realized that the PRS experiment is the one of type, in which a small effect of a physical difference of two neighboring $\phi$- worlds is measured by means of an information exchange. The effect is of the second order *under weak field conditions.* It is determined by a potential difference (24) between an emitter and a detector, placed on different equipotential surfaces. *The effect is not affected by the state of free fall.* In fact, this is the test of the $\phi$-world as compared to the asymptotically flat space, which has the absolute potential $\phi = 0$.

The concept of asymptotically flat space, in which a free particle has a maximal proper mass, leads to the fundamental physical problem of the mass origin. Obviously, it relates to cosmology and cannot be resolved in the Standard Cosmological Model. In this respect one has to appreciate the GRT approach to the energy conservation problem making ends somehow meet in an asymptotically flat space at space infinity. However, the idea of a limited universe evolving in an infinite "free" space does not look promising in many respects. For example, it does not allow considering a force propagator exchange with a "universe matter" for resolving the Mach principle problem. A multi-universe concept advocated, for example, by Martin Rees [9] has more physical sense.

Astrophysics undoubtedly will benefit from a new development of gravitational theory. The GRT "black hole" physical concept, in particularly, could be essentially reconsidered. From the SRT standpoint light could not be trapped in principle, and the paradox of infinite time of a black hole creation would not arise.

*Conclusion*

On the basis of the relativistic mass-energy concept we found that a proper mass of a test particle in a gravitational field depends on a potential energy, hence, a freely falling particle has a varying proper mass. Consequently, a multitude of freely falling reference frames cannot be regarded as the multitude of equivalent inertial reference frames. On the other hand, the classical Galilean-Newtonian concept of equivalent inertial frames has been successfully generalized in the SRT. For this reason one has to doubt whether the EP could be consistent with the relativistic theory of gravity.

The suggested test on the EP violation is crucial. In combination with the original PRS (stationary) experiment a new frequency shift measurement in a freely falling frame will allow discriminating between two different approaches to the field theory. The question to be answered is whether a photon gravitates. It is shown that existing observational data are not sufficient to make a final conclusion in this respect, and the suggested test is the one on falsifying the EP. Luckily, we have the Mossbauer instrumentation needed for further probing a gravitational field with a photon with a precision adequate to the test requirements.